\documentclass{aa}

\usepackage{txfonts}
\usepackage{graphicx}
\usepackage{astrobib}

\newcommand{\gcn}{GCN Circular}

\begin{document}

%%%%%%%%%%%%%%%%%%%%%%%%%%%%%%%%%%%%%%%%%%%%%%%%%%%%%%%%%%

\title{Robust photometric redshift determinations of gamma-ray burst afterglows at $z \gtrsim 2$}
%\subtitle{}

\titlerunning{Robust photometric redshift determinations of GRB afterglows at $z \gtrsim 2$}

  \author{
    P.A.~Curran\inst{1}
    \and R.A.M.J.~Wijers\inst{1}
    \and M.H.M.~Heemskerk\inst{1}
    \and R.L.C.~Starling\inst{2}
    \and K.~Wiersema\inst{2}
    \and A.J.~van~der~Horst\inst{3}
  }

%\authorrunning{P.A.~Curran et al.}

  \offprints{P.A.~Curran (pcurran@science.uva.nl)}
  
  \institute{
    Astronomical Institute, University of Amsterdam, Kruislaan 403, 1098\,SJ Amsterdam, The Netherlands 
    \and Department of Physics and Astronomy, University of Leicester, University Road, Leicester LE1 7RH, UK 
    \and NASA Postdoctoral Program Fellow, NSSTC, 320 Sparkman Drive, Huntsville, AL 35805, USA
  }

\date{Received ; accepted}

%%**************************************************
%% Abstract
%%**************************************************

\abstract
% context (optional)
{Theory suggests that about 10\% of \emph{Swift}-detected gamma-ray bursts (GRBs) will originate at redshifts, $z$, greater than 5 yet a number of high redshift candidates may be left unconfirmed due to the lack of measured redshifts.}
% aims(mandatory)
{Here we introduce our code, \emph{GRBz}, a method of simultaneous multi-parameter fitting of GRB afterglow optical and near infrared, spectral energy distributions. It allows for early determinations of the photometric redshift, spectral index and host extinction to be made.} 
% methods(mandatory)
{We assume that GRB afterglow spectra are well represented by a power-law decay and model the effects of absorption due to the Lyman forest and host extinction. We use a genetic algorithm-based routine to simultaneously fit the parameters of interest, and a Monte Carlo error analysis.} 
% results (mandatory)
{We use GRBs of previously determined spectroscopic redshifts to prove our method, while also introducing new near infrared data of \object{GRB\,990510} which further constrains the value of the host extinction.}
% conclusion (optional)
{Our method is effective in estimating the photometric redshift of GRBs, relatively unbiased by assumptions of the afterglow spectral index or the host galaxy extinction. 
Monte Carlo error analysis is required as the method of error estimate based on the optimum  population of the genetic algorithm underestimates errors significantly. }


\maketitle

%%**************************************************
%% Section 1: Introduction
%%**************************************************
\section{Introduction}\label{grbz:intro}

Theory suggests and observations agree that approximately 10\% of \emph{Swift}-detected gamma-ray bursts (GRBs) will originate at redshifts, $z \gtrsim 5$ \cite{bromm2006:ApJ642,jakobsson2006:A&A447} and about 3\% at redshifts, $z \gtrsim 6$ \cite{daigne2006:MNRAS372}. Furthermore, due to the favourable $K$-correction at a fixed observer time and cosmological time dilation, the observed optical flux is not expected to fade significantly with increasing redshift. Hence GRBs should be detectable out to $z \gtrsim 10$ \cite{ciardi2000:ApJ540,lamb2000:ApJ536}, making them the most distant observable objects in the Universe. Yet a number of high redshift candidates may be left unconfirmed due to the lack of observed counterparts or lack of observations \cite{ruiz2007:ApJ669}.

 The most usual way of determining the redshift of a GRB is to spectroscopically measure the redshift of the afterglow at early times if it is bright enough, or to measure that of its host galaxy, either photometrically (e.g. \citeNP{bolzonella2000:A&A363}) or spectroscopically, once the optical afterglow of the burst has dimmed sufficiently. Neither of these may be possible given the dimness of the afterglows \cite{fynbo2007:Msngr130} and the host galaxies.
In addition, programs to obtain such measurements are frequently only triggered if there is evidence to suggest a high redshift, or a burst of significant interest. 
There are a number of redshift indicators, or pseudo redshift methods, which rely on correlations between observable properties and redshift (e.g., \citeNP{guidorzi2005:MNRAS364,amati2006:MNRAS372,pelangeon2006:GCN5004}) that offer approximate redshifts to varying degrees of success.

An alternative method of redshift determination, useful when the source is not bright enough for spectroscopic observations, is to photometrically measure the redshift of the afterglow itself, given enough simultaneous optical/near infrared data points.  Attempts at photometric redshift determination usually assume a spectral index and a value for host extinction, while fitting for redshift, or fit for a limited number of discrete values of spectral index and host extinction (e.g., \citeNP{andersen2000:A&A364,jakobsson2006:A&A447}).  
Independently, many attempts have been made to fit host extinction and spectral index to similar data, when the redshift is previously known (e.g., \citeNP{galama2001:ApJ549,stratta2004:ApJ608,kann2006:ApJ641,schady2007:MNRAS377,starling2007:ApJ661}). This is important as the properties of the circumburst medium and host galaxy can be used to constrain progenitor types and burst models themselves.

In this paper we present our code, \emph{GRBz}, which, unlike other methods of photometric redshift determination, fits all three parameters simultaneously and allows for determinations of the photometric redshift to be made, relatively unbiased by assumptions of the spectral index or the host extinction. 
In section \ref{grbz:method} we describe the method of modelling and fitting that we apply to the data described in section \ref{grbz:data}, including our previously unpublished near IR data of GRB\,990510. In section \ref{grbz:results} we present our results and in section \ref{grbz:discussion} we discuss the results as they apply to the fitting mechanism. We summarise our findings in section \ref{grbz:conclusions}.

%%**************************************************
%% Section 2:
%%**************************************************

\section{Method}\label{grbz:method}
  
\subsection{Model}\label{grbz:model}

Our method is based on calculating the flux at a frequency, $F(\nu)$, for a given set of fit parameters: redshift, $z$, spectral index of afterglow, $\beta$ and host extinction, $E_{B-V}$. This is done by assuming that the emitted flux from the afterglow is well represented by a power-law decay ($F(\nu) \propto \nu^{-\beta}$), and adjusting this for extinction in the host galaxy, line blanketing associated with the Lyman forest in the intervening medium and extinction in the Milky Way. This flux is then integrated, numerically, over the transmission curve of an individual filter, $T_{\mathrm{filter}}(\nu)$: $\nu_{\mathrm{min}} \leq \nu \leq \nu_{\mathrm{max}}$, and normalised to give the flux as measured by that filter,
  \[ 
  F_{\mathrm{filter}} =  \frac{\int_{\nu_{\mathrm{min}}}^{\nu_\mathrm{max}} F(\nu) T_{\mathrm{filter}}(\nu) d\nu}{\Delta\nu} , 
  \]
  where the effective width of the filter is
\[
\Delta\nu = \int_{\nu_{\mathrm{min}}}^{\nu_{\mathrm{max}}}  T_{\mathrm{filter}}(\nu)    d\nu . 
\] 
This is repeated for each of the filters in the Spectral Energy Distribution (SED) so that the calculated fluxes may be compared to those measured, by a Chi Squared ($\chi^2$) test, and best fit parameters estimated.

The extinction in the host galaxy and the Milky Way are calculated using the models of \citeN{pei1992:ApJ395} or \citeN{calzetti2000:ApJ533}. We use the Pei model for the MW, SMC or LMC, or  the Calzetti model for a star forming region, with the corresponding values of  $R_{V} = A_{V}/E_{B-V}$ \cite{cardelli1989:ApJ345} from those authors,  as an approximation of a GRB host galaxy. Note that here we use $E_{B-V}$ to parametrise the extinction -- not $A_{V}$ as is often used -- as it follows naturally from Pei's treatment and simplifies the calculation of extinction at a given frequency, $A_{\nu}$.

  The Lyman absorption, by neutral hydrogen in the intergalactic medium, is calculated from the model presented by \citeN{madau1995:ApJ441}. Though Madau only gives the first 4 coefficients, $A_{j}$ (Lyman $\alpha, \beta, \gamma, \delta$) of the model, the remaining 13 ($\epsilon...\rho$) may be extrapolated from those in \citeN{madau1996:MNRAS283} due to their linear relationship with the wavelengths from the Balmer formula for the Lyman series: 
% \begin{equation}
\[
\frac{1}{\lambda} = \left( 1 -\frac{1}{\mathrm{n}^{2}} \right) \mathbb{R}, 
\]
% \end{equation}
where n = 2, 3, ... , 18 and $\mathbb{R}$ is the Rydberg constant.
This does not take into account the effect of a damped Lyman alpha absorber, which may or may not be present in the host galaxy. If present this would cause the flux to be under corrected and hence cause an overestimate of the redshift.

 As the transmission curves of the filters are read into the fitting program directly and the necessary effective frequencies and widths calculated, any filter can be used. We have initially used the Johnson $UBVRIK$ \cite{johnson1965:ApJ141},  2MASS $JHK_{S}$ \cite{cohen2003:AJ126} and \emph{Swift} UVOT filters \cite{roming2005:SSRv120}, covering the approximate frequency range $10^{14} - 10^{15}$\,Hz.

Independently of the fitting mechanism, it is useful to be able to approximate the relative effect that Lyman absorption, $T_{\mathrm{Lyman}}(\nu, z)$, will have on observed flux; our code also allows us to calculate an effective transmission of a filter at a given redshift, $z$, and for a given spectral index, $\beta$:
\[
T_{\mathrm{effective Lyman}}(z, \beta, T_{\mathrm{filter}}) =  \frac{\int_{\nu_{\mathrm{min}}}^{\nu_{\mathrm{max}}} T_{\mathrm{filter}}(\nu)  T_{\mathrm{Lyman}}(\nu, z) \nu^{-\beta} d\nu}{\int_{\nu_{\mathrm{min}}}^{\nu_{\mathrm{max}}}  T_{\mathrm{filter}}(\nu)  \nu^{-\beta} d\nu} . 
\]
This gives a correction factor by which to divide the observed flux so as to estimate the flux as if unaffected by Lyman absorption, as successfully implemented in \citeN{starling2007:ApJ661}.

\subsection{Fitting}\label{grbz:fitting}

The fitting method implemented within our C-program is the genetic algorithm-based optimisation subroutine {\tt PIKAIA} \cite{charbonneau1995:ApJS101}, which we use to minimise the $\chi^{2}$ of the data points. Due to the complexity, and many local minima in our solution space, this proved to be a much more robust, stable and reproducible method than those based on \emph{simulated annealing} or \emph{downhill simplex} methods (section 10.9 of \citeNP{press1992:nrca.book} and references therein and \citeNP{nelder1965:CompJ.7.308}, respectively).
This method allows us to fix, or constrain, the possible values of solutions -- for example, if a spectroscopic redshift has been determined or the spectral index has been constrained from a temporal index -- while not being overly dependent on the starting values. (For an example of the previous use of the genetic algorithm in astronomy see \citeNP{mokiem2005:A&A441}).

GRB afterglows can be faint and fade rapidly, so often only upper limits on the afterglow brightness  are available and we require a standard procedure for dealing with this. 
In these cases we take the flux to be zero and use the limit as an estimate of error. We have compared this method to a method where the $\chi^{2}$ is defined as zero when the model flux is within the limit and very large elsewhere, and find that the methods give consistent results.

%\subsection{{\bf error analysis}}

Uncertainties of the fit parameters are estimated via a Monte Carlo analysis, whereby the data are randomly perturbed within the Gaussian distribution of their errors, and refit multiple times. Due to the slowness of these calculations caused by the need to numerically integrate over the filters multiple times we choose a modest number ($\sim 10^{2}$) of trials. The average of these perturbed fits should agree with the initial unperturbed fit and the distribution of the parameters should give a accurate approximation of the uncertainties.  
From the frequency distribution of any one parameter, we see that the distribution is clearly not Gaussian so the usual method of quoting sigma errors is not valid, though we do quote the uncertainties as nominal, symmetric 68\% confidence intervals. 
A population analysis method may also be used to approximate errors directly from the fitting algorithm \cite{mokiem2005:A&A441} but the Monte Carlo method is more robust. See also Figure\,\ref{monte-plot} and section \ref{grbz:errors}.

%%**************************************************
%% Section 3:
%%**************************************************

\section{Data}\label{grbz:data}

\subsection{Sample selection}

\begin{table}	
  \centering	
  \caption{Optical SED references for the GRBs in our sample.} 	
  \label{optical_table} 	
  \begin{tabular}{l l l } 
    \hline\hline
    GRB & Filters & Reference\\ 
    \hline 
    990510	& $K_{S},H,J,I,R,V,B$       &  1\\ 
    000131	& $K,H,I,R,V$              &  2 \\ 
    050319	& $V,B,U,UVW1,UVM2,UVW2$                     &  3\\ 
    050814	& $K,J,I,R,V$       &  4 \\ 
    050904	& $K,H,J,I,R,V$	                   &   5 \\ 
    \hline 
  \end{tabular}
  \begin{list}{}{}
  \item[]
    1 \citeN{stanek1999:ApJ522}; \citeN{holland2000:A&A364}; this article,
    2 \citeN{andersen2000:A&A364},
    3 \citeN{mason2006:ApJ639},
    4 \citeN{jakobsson2006:A&A447},
    5 \citeN{tagliaferri2005:A&A443}
  \end{list}
\end{table}

The purpose of this paper is to introduce our  method of photometric redshift determination, not to provide a statistical analysis of a large number of bursts, so  we have ``cherry-picked'' a small sample of bursts (Table\,\ref{optical_table}).  GRBs were selected from the literature on the basis of availability of an SED and, knowing that the Lyman forest would only affect optical bands at $z \gtrsim 2.5$, redshift.
Bursts with previously determined properties, especially a spectroscopic redshift, that could be compared to our fit parameters were preferred though in the case of GRB\,050814 only a photometric redshift was available. The previously published optical data for GRB\,990510 were augmented by our near infrared (nIR) observations of the source.

\subsection{Near IR observations of GRB\,990510}\label{grbz:data_990510}

From May 10 to May 22, 1999, at 4 epochs, a total of 186 60-second exposures, or frames, in $J$, $H$ and $K_{S}$ filters, were obtained of the optical afterglow of GRB\,990510.
The data were obtained with the Son of ISAAC (SofI) infrared spectrograph and imaging camera on the 3.58m ESO-New Technology Telescope (NTT). The NTT-SofI data were reduced using the {\small IRAF} package wherein flatfielding, sky subtraction and frame addition were carried out. 
Relative PSF photometry was carried out on the final images using the {\small DAOPHOT} package (Stetson 1987) within {\small IRAF}. The PSF model was created using 16 stars in the field, one of which was a 2MASS object suitable for calibration (2MASS designation: 13380057-8029119, RA Dec (J2000): 13:38:00.58 -80:29:11.9). The resultant magnitudes and $1 \sigma$ errors of the nIR counterpart are shown in Table \ref{obs_table990510}.

%%%%%%%%%%%%%%%%%%%%%% TABLE %%%%%%%%%%%%%%%%%%%%%%%%%%%%%%

\begin{table}
  \begin{center}
    \caption{Near IR Observations of GRB\,990510. The magnitudes are uncorrected for the Galactic extinction, $E_{B-V} = 0.203$.}
    \label{obs_table990510}
    \begin{tabular}{l c c c l}
      \hline\hline
      $T_{\mathrm{mid}}$ & Seeing    & Band  & $N_{\mathrm{exp}} \times T_{\mathrm{exp}}$ & Magnitude \\
      days    & $^{\prime\prime}$      &       & $\times$ sec    &    \\
      \hline
      0.601     & 0.68      & $J$       & 10$\times$60 & 17.51 $\pm$ 0.07\\
      1.000     & 1.00      & $J$       & 10$\times$60 & 18.16 $\pm$ 0.08\\
      3.837     & 1.44      & $J$       & 26$\times$60 & 20.66 $\pm$ 0.24\\
      0.610     & 0.62      & $H$       & 10$\times$60 & 16.96 $\pm$ 0.08\\
      1.014     & 1.26      & $H$       & 10$\times$60 & 17.52 $\pm$ 0.09\\
      3.862     & 1.32      & $H$       & 20$\times$60 & 19.92 $\pm$ 0.28\\
      11.815    & 0.65      & $H$       & 60$\times$60 & 22.10 $\pm$ 0.34\\
      0.618     & 0.63      & $K_{\mathrm{S}}$   & 10$\times$60 & 16.25 $\pm$ 0.17\\
      1.024     & 1.06      & $K_{\mathrm{S}}$   & 10$\times$60 & 16.85 $\pm$ 0.18\\  
      3.881     & 1.34      & $K_{\mathrm{S}}$   & 20$\times$60 & 19.25 $\pm$ 0.25\\
      \hline
    \end{tabular}
  \end{center}
\end{table}

%%%%%%%%%%%%%%%%%%%%%%%%%%%%%%%%%%%%%%%%%%%%%%%%%%%%%%%%%%

 \begin{figure} 
  \centering 
  \resizebox{85mm}{!}{\includegraphics[angle=-90]{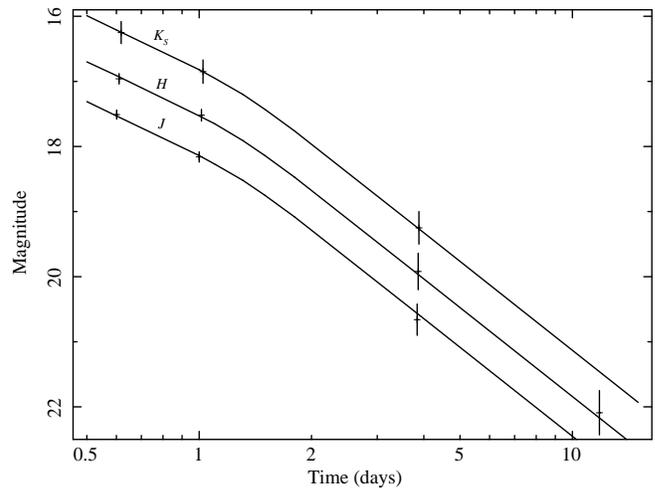}}
  \caption{Near IR light curve of GRB\,990510 with the simultaneous fit shown.} 
  \label{990510_lc} 
\end{figure}

%%%%%%%%%%%%%%%%%%%%%%%%%%%%%%%%%%%%%%%%%%%%%%%%%%%%%%%%%%

There is clearly a break to a steeper slope between our second and third epochs of observation (Figure\,\ref{990510_lc}). This is consistent with the measurement of a jet break at 1.31 days after the trigger time \cite{zeh2006:ApJ637}. Using the simultaneous fit method outlined in \citeN{curran2007:A&A467} and this break time we find that the temporal indices are  $\alpha_{1} = 1.06 \pm 0.20$ and $\alpha_{2} = 1.82 \pm 0.18$ ($1 \sigma$ uncertainties). These values differ from those found by \citeN{zeh2006:ApJ637}, most likely because of the lack of temporal sampling of the nIR light curve.

The nIR magnitudes of GRB\,990510 were interpolated to a common time ($t = 0.61$\,days) and then converted to flux values using the calibration of  \citeN{cohen2003:AJ126}. $VRI$ data at the same epoch were calculated from the light curve fit of \citeN{holland2000:A&A364}, while $B$ data was calculated from the light curve of \citeN{stanek1999:ApJ522}. All data were then corrected for Galactic extinction of $E_{B-V} = 0.203$ \cite{schlegel1998:ApJ500} at their various frequencies \cite{pei1992:ApJ395}.

%%**************************************************
%% Section 4
%%**************************************************

\section{Results}\label{grbz:results}

For all the fits presented here we have assumed that the host galaxy extinction is well represented by the \citeN{pei1992:ApJ395} SMC model. This is supported by studies of GRB host extinction \cite{galama2001:ApJ549,stratta2004:ApJ608,kann2006:ApJ641,schady2007:MNRAS377,starling2007:ApJ661}, though it is not necessarily the model which best fits each of these individual bursts
(e.g., GRB\,050904; \citeNP{stratta2007:ApJ661}).
As explained in section \ref{grbz:fitting}, best fit parameters are given as the average values of those returned from a Monte Carlo analysis (Table\,\ref{result-table}), uncertainties are quoted as symmetric 68\% confidence intervals. Figure \ref{grbz:seds-1} shows the SEDs for each of the bursts in our sample. The flux is calculated from the best fit parameters and clearly shows the cut off due to the line blanketing associated with the Lyman forest.

\begin{figure}[!]
  \centering 
  \resizebox{85mm}{!}{\includegraphics[angle=-90]{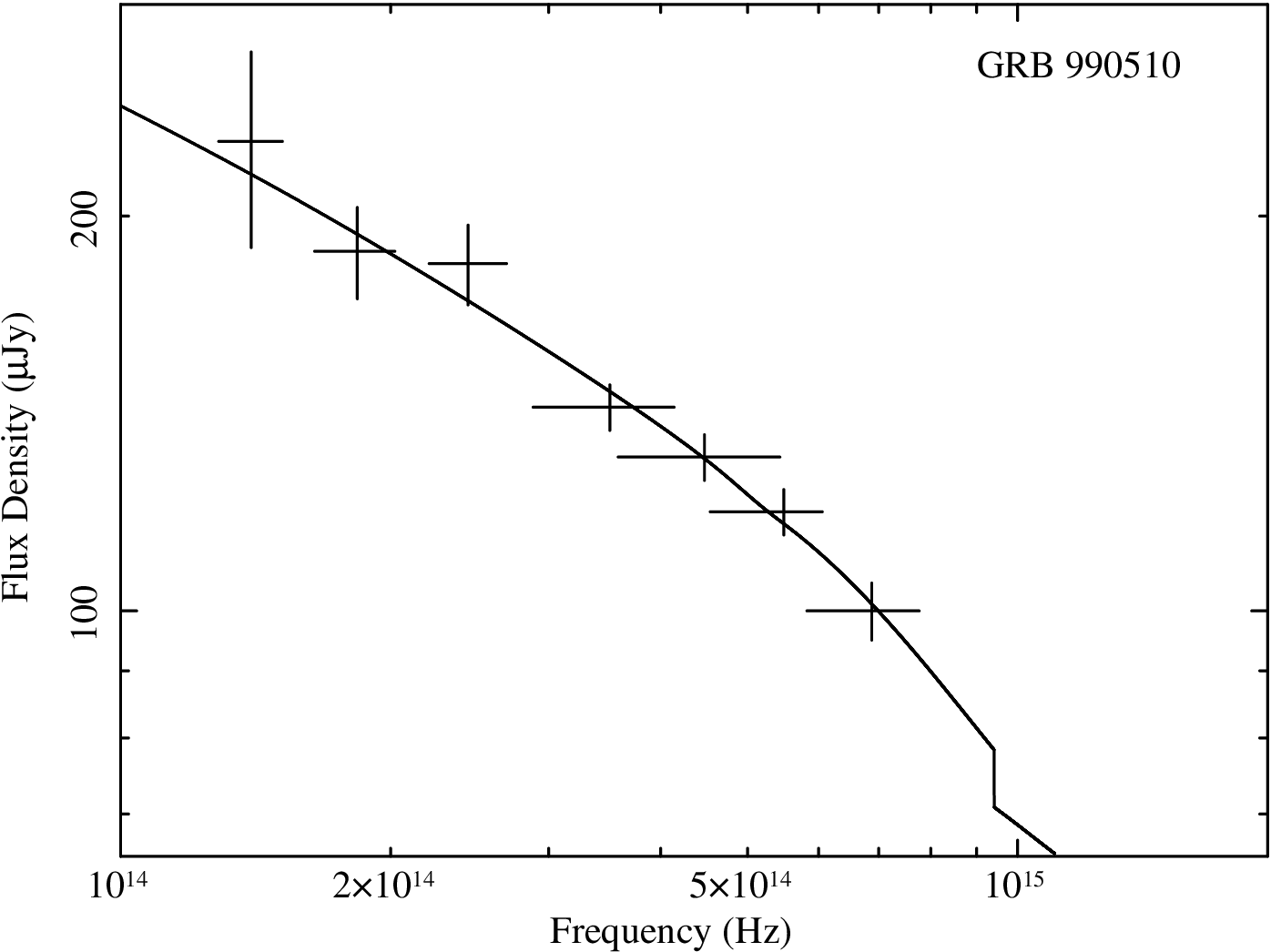} }
  \vspace{2mm}\vspace{2mm}
  \resizebox{85mm}{!}{\includegraphics[angle=-90]{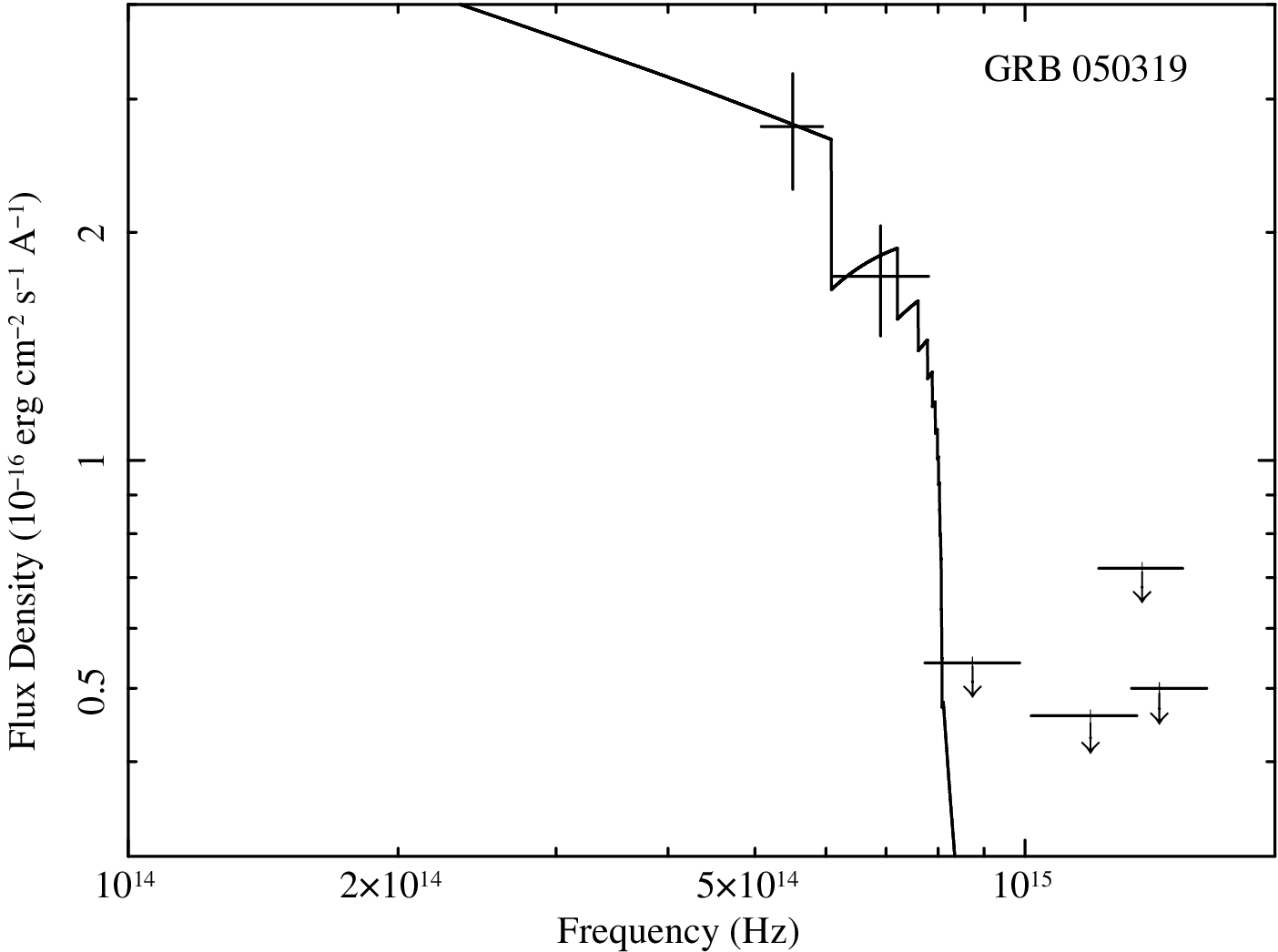} }
  %\resizebox{85mm}{!}{\includegraphics[angle=-90]{grb000131.ps} }
  \vspace{2mm}\vspace{2mm}
  \resizebox{85mm}{!}{\includegraphics[angle=-90]{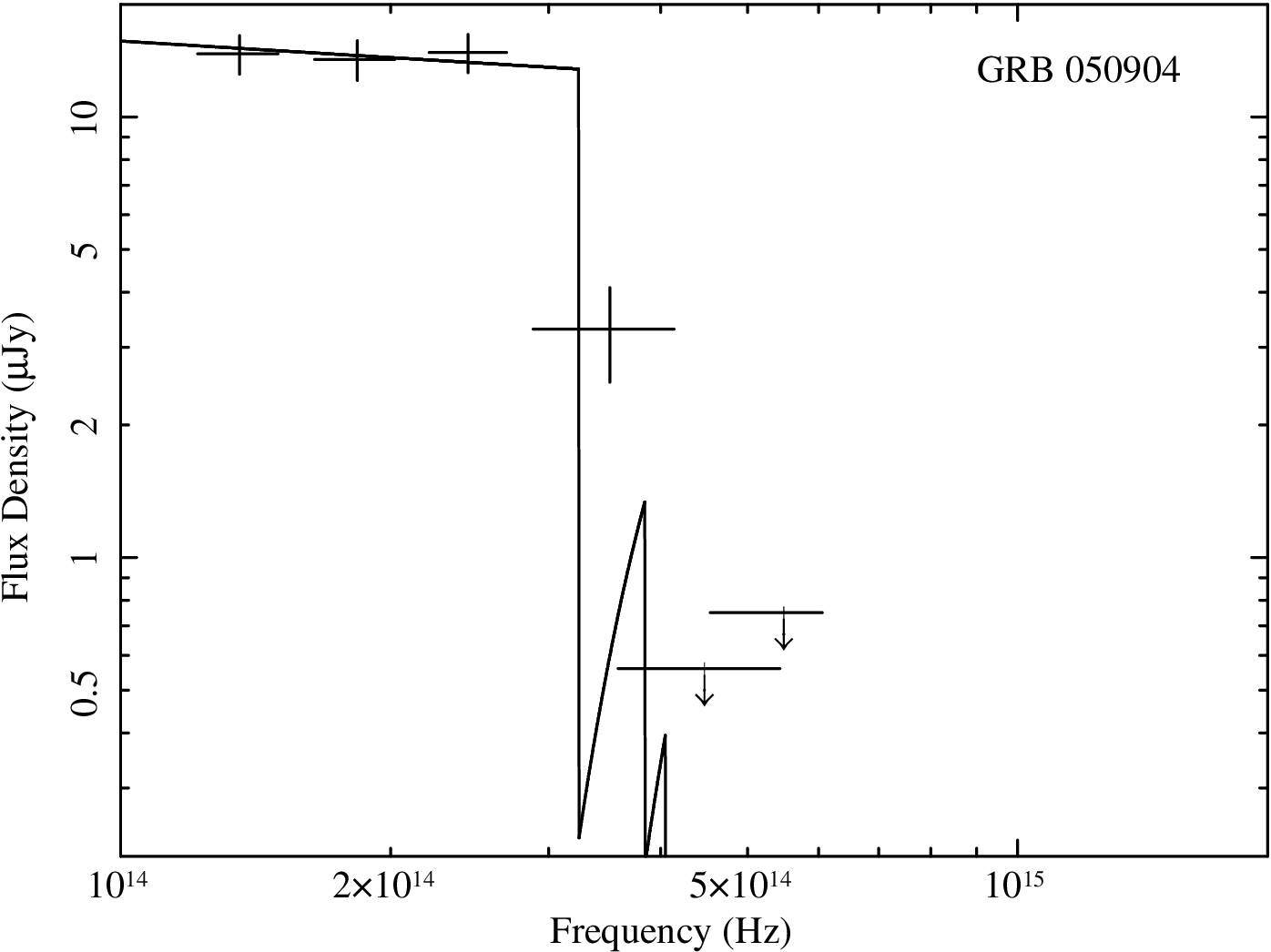} }
  % \resizebox{85mm}{!}{\includegraphics[angle=-90]{grb050319.ps} }
  \caption{Fits to the SEDs of our sample GRBs (for filters used in each SED see Table\,\ref{optical_table}). The errors on the frequency axis represent the width at Transmission = 0.2.}  
  \label{grbz:seds-1} 
\end{figure} 

\setcounter{figure}{1}

\begin{figure}[!]
  \centering 
  \resizebox{85mm}{!}{\includegraphics[angle=-90]{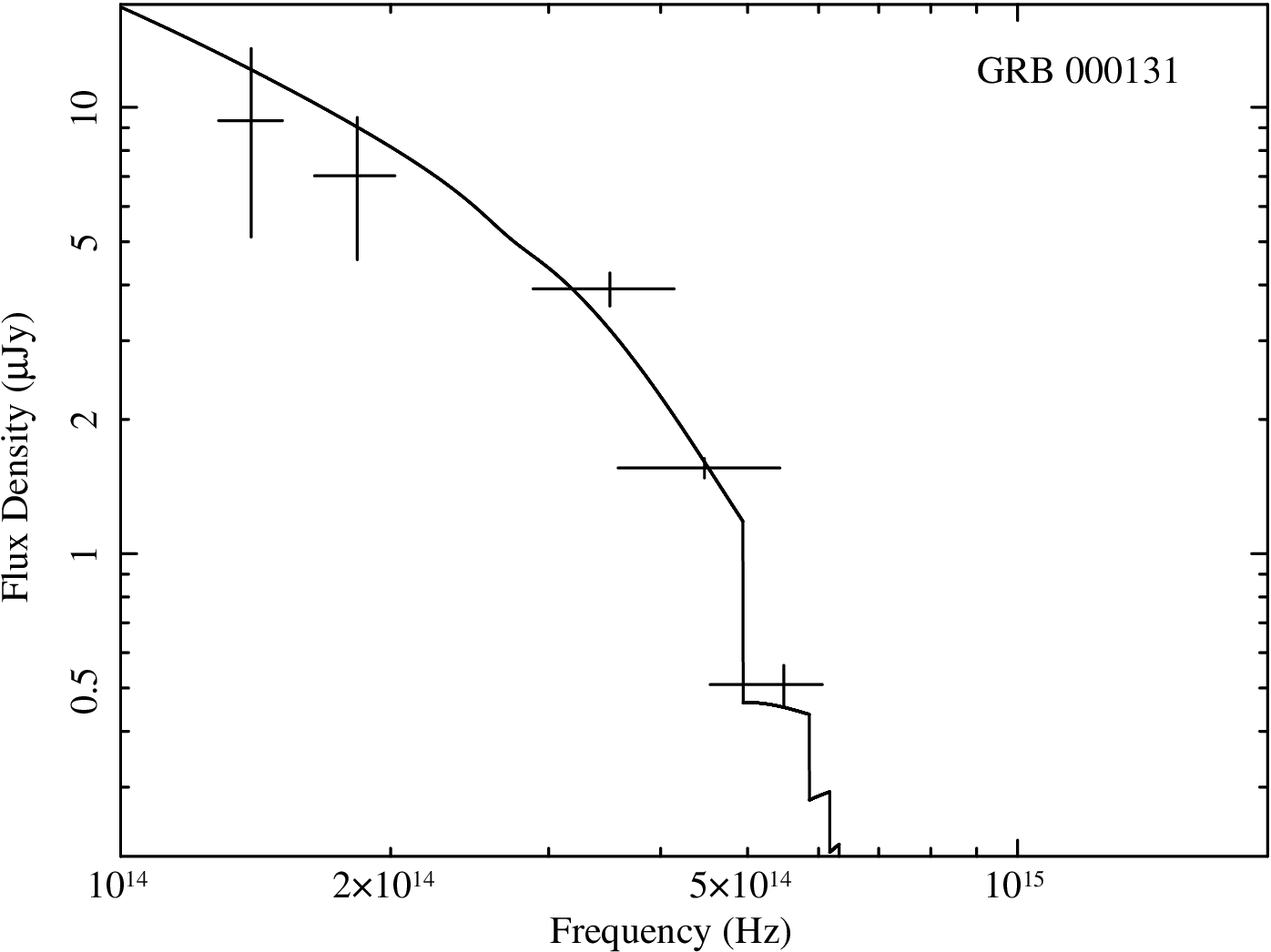} }
  % \resizebox{85mm}{!}{\includegraphics[angle=-90]{grb050814.ps} }
  \vspace{2mm}\vspace{2mm}
  \resizebox{85mm}{!}{\includegraphics[angle=-90]{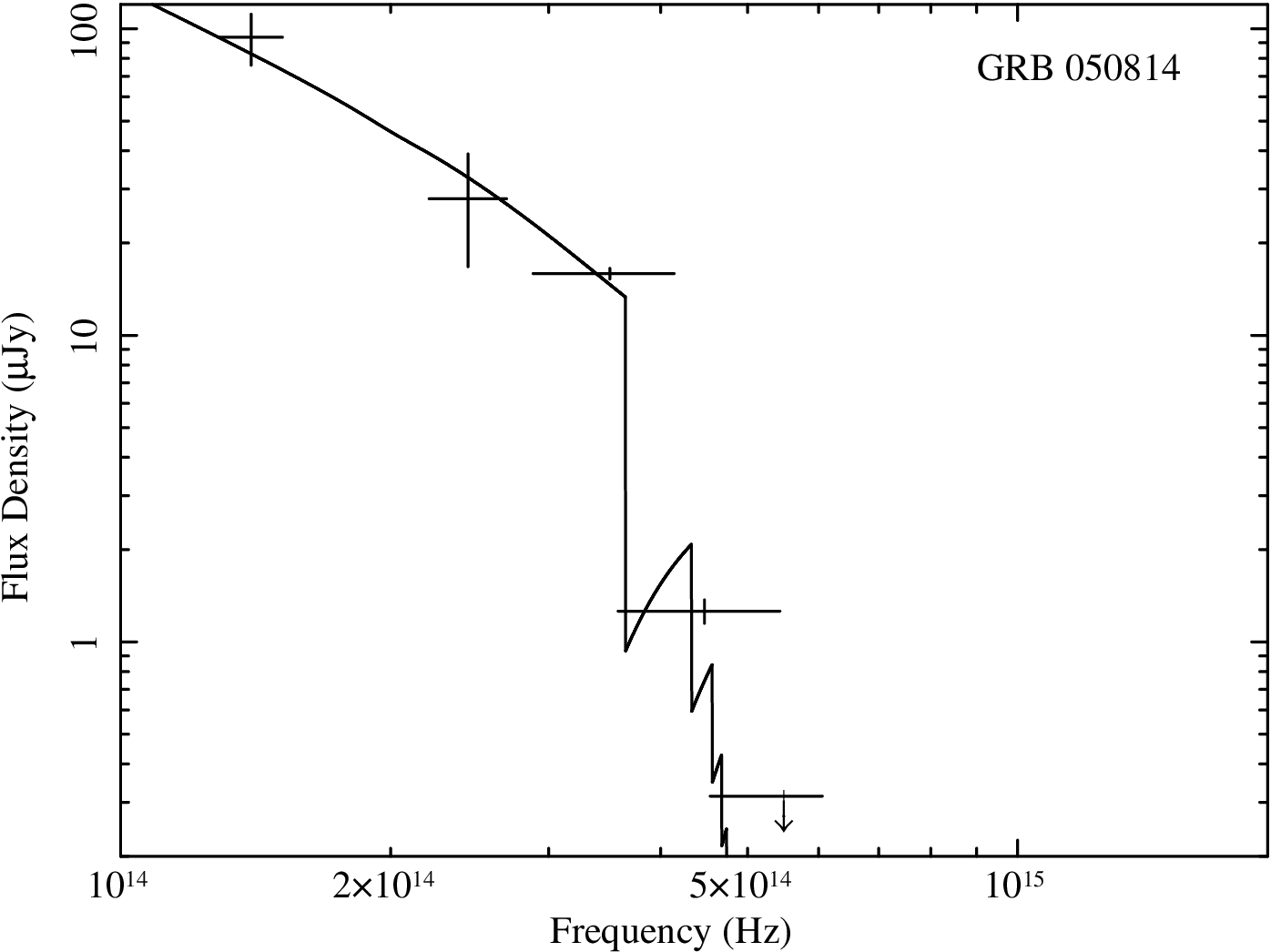} }
  % \resizebox{85mm}{!}{\includegraphics[angle=-90]{grb050904.ps} }
  \caption{Continued}
  \label{grbz:seds-2} 
\end{figure}

\subsection{GRB\,990510}\label{grbz:990510}

GRB\,990510 has a known spectroscopic redshift, $z_{{\rm spec}} = 1.619 \pm 0.002$ \cite{vreeswijk1999:GCN324,vreeswijk2001:ApJ546} which is below the operational limits of our program, hence we constrain the redshift in our fit to be within the errors of the known value while leaving the other parameters free. We find that $\beta = 0.35\pm 0.17$ and $E_{B-V} < 0.05$, corresponding to a rest frame extinction, $A_{V} < 0.14$. 
This spectral index is lower than, but consistent with, the value found by \citeN{starling2007:ApJ661} of $0.53^{+0.07}_{-0.01}$ who utilised X-ray measurements as well as our nIR data. 
Our estimate of host extinction is consistent with that of \citeANP{starling2007:ApJ661} and \citeN{stratta2004:ApJ608} who also utilised X-ray measurements; all of which indicate negligible extinction, though \citeANP{starling2007:ApJ661}'s approximation is much more tightly constrained than ours.

 \subsection{\object{GRB\,000131}}

While GRB\,000131 has a spectroscopic redshift of $z_{{\rm spec}} = 4.500 \pm 0.015$ \cite{andersen2000:A&A364}, we have left it, as well as the other parameters in our fit, free.
From \citeANP{andersen2000:A&A364}'s published SED, we have found that $z = 4.2 \pm 0.4$,  $\beta = 0.7 \pm 0.4$ and $E_{B-V} = 0.10 \pm 0.06$ ($A_{V} = 0.29 \pm 0.18$).  This extinction is consistent with that found by \citeANP{andersen2000:A&A364} who assumed a spectral slope, $\beta = 0.70$ based on the constraints imposed by the temporal decay index.

 \subsection{\object{GRB\,050319}}

The UVOT telescope on \emph{Swift} observed this burst in the UV and optical at early times ($T_0$ + 240--290\,s).  Using the published SED  \cite{mason2006:ApJ639} we find that $z = 3.0 \pm0.2 $,  $\beta = 0.6 \pm 0.3$ and $E_{B-V} < 0.07$  ($A_{V} < 0.20$). \citeN{jakobsson2006:A&A460} determine a spectroscopic redshift of  $z_{{\rm spec}} = 3.24$ consistent with our result,  while \citeN{mason2006:ApJ639} infer a spectral index of $\beta = 0.8 \pm 0.1$ from the X-ray light curve. Though our spectral index is consistent with this, we cannot infer much information about the spectral index or extinction since there is only one data point above the Lyman break.

 \subsection{\object{GRB\,050814}}

\citeN{jakobsson2006:A&A447} find a photometric redshift, $z = 5.3 \pm 0.3$, and fixing $\beta = 1.0$ find an extinction of A$_{V} = 0.9$. Using their data points we find a redshift of $z = 5.77 \pm 0.12$. Since there are only 2 data points above the break, the spectral index and extinction may only be loosely constrained as $\beta = 0.9 \pm 0.5$ and $E_{B-V} = 0.08 \pm 0.05$ ($A_{V} = 0.23$). 
This spectral index is consistent with that of the X-ray index, $\beta_{\mathrm{X}} = 0.8 \pm 0.2$ \cite{morris2005:GCN3805}. 
\citeANP{jakobsson2006:A&A447} point out that their extinction is marginally higher than inferred from other bursts with bright optical counterparts \cite{kann2006:ApJ641}. While a low extinction is not necessarily the case, our result is in line with those that \citeANP{jakobsson2006:A&A447} compare theirs to. These authors also point out that their extinction and spectral index would overestimate, slightly, the X-ray flux at the time. The alternative spectral index they suggest to compensate for this ($\beta = 1.1$) is within the uncertainties of our estimate.

%%%%%%%%%%%%%%%%%%%%%% TABLE %%%%%%%%%%%%%%%%%%%%%%%%%%%%%%
\begin{table*}[!]
  \begin{center}
    \caption{Results of our simultaneous fits and previously published, spectroscopic or photometric, redshifts, $z_{{\rm literature}}$, for comparison.}
    \label{result-table}
    \begin{tabular}{l l l l l}
      \hline\hline
      GRB        & $z_{{\rm literature}}$ & $z$  & $E_{B-V}$     & $\beta$\\
      \hline      
      
      990510     &  1.619$\pm 0.002$$^{\mathrm{a}}$   &	             &  $<$ 0.05          &  0.35 $\pm$ 0.17 \\
      000131     &  4.500$\pm 0.015$$^{\mathrm{b}}$   & 4.2 $\pm$ 0.4 &  0.10 $\pm$ 0.06   &  0.7 $\pm$ 0.4 \\
      050319     &  3.240$\pm 0.001$$^{\mathrm{c}}$               & 3.0 $\pm$ 0.2 &  $<$ 0.07          &  0.6 $\pm$ 0.3 \\
      050814     &  5.3$\pm$0.3$^{\mathrm{d}}$        & 5.77 $\pm$ 0.12 &  0.08 $\pm$ 0.05 &  0.9 $\pm$ 0.5 \\
      050904     &  6.30$\pm 0.002$$^{\mathrm{e}}$    & 6.61 $\pm$ 0.14 & $<$ 0.03         &  0.05 $\pm$ 0.02 \\
      \hline
    \end{tabular}
    \begin{list}{}{}
    \item[]$^{\mathrm{a}}$ \citeN{vreeswijk1999:GCN324}, $^{\mathrm{b}}$ \citeN{andersen2000:A&A364},
      $^{\mathrm{c}}$ \citeN{jakobsson2006:A&A460}, 
      $^{\mathrm{d}}$ \citeN{jakobsson2006:A&A447}, $^{\mathrm{e}}$ \citeN{kawai2006:Natur440}. 
    \end{list}
  \end{center}
\end{table*}

%%%%%%%%%%%%%%%%%%%%%%%%%%%%%%%%%%%%%%%%%%%%%%%%%%%%%%%%%%%%%%%%%%%%

 \subsection{\object{GRB\,050904}}

The highest redshift GRB identified to date has been the spectroscopically measured, $z = 6.295 \pm 0.002$ \cite{kawai2006:Natur440}, GRB\,050904. Using the data published by \citeN{tagliaferri2005:A&A443}, we find a redshift of $z = 6.61 \pm 0.14$ which is consistent at the $2\sigma$ level with the spectroscopic value.  Since the nIR flux points, unaffected by the break, are all consistent with a flat spectrum we can infer very little from the obtained values of $\beta = 0.05 \pm 0.02$ and $E_{B-V} \lesssim 0.03$.

%%**************************************************
%% Section 5: Discussion
%% **************************************************

\section{Discussion}\label{grbz:discussion}

\subsection{Limitations}\label{grbz:limitations}

\begin{figure} 
  \centering 
  \resizebox{85mm}{!}{\includegraphics[angle=-90]{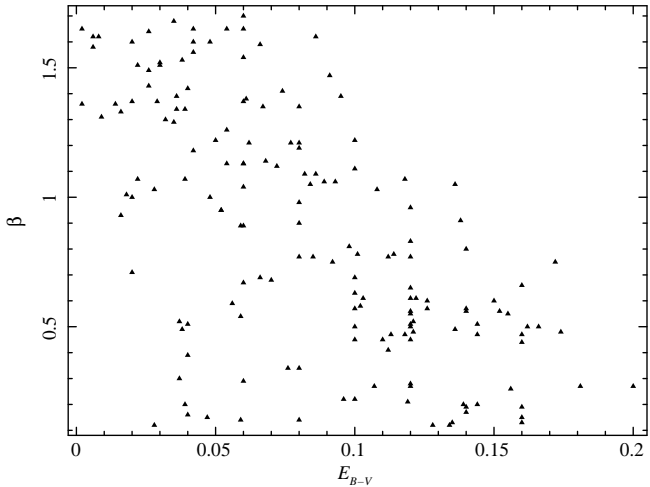}}
  \caption{Spectral index, $\beta$, versus host extinction, $E_{B-V}$, for the Monte Carlo analysis of GRB\,050814.}  
  \label{050814_scatter} 
\end{figure}

We are able to draw a number of general conclusions from the data sets that we have analysed here, together with synthesised data sets that we initially tested the fitting method on, regarding the operational limits of the program.  Obviously, whether or not any one of the three parameters with physical meaning ($z$, $\beta$ and $E_{B-V}$) can be fit with accuracy is dependent on the data available. 
We may only fit the redshifts of bursts with $z \gtrsim 2.5$ (lower if UVOT data are available) as this is where the Lyman forest starts to affect the high frequency filters ($UB$). Redshifts lower than this act degenerately, so a fit that has a wide spread of redshifts under, or around 2.5 must be assumed to be under this limit and can only be fit, in the other parameters, if the redshift is previously known (e.g. GRB\,990510).

The accurate determination of $\beta$, $E_{B-V}$ or host extinction model is dependent on the number of data points above the Lyman break. If there are few data points unaffected by Lyman absorption we cannot expect to draw any firm conclusions regarding $\beta$ or $E_{B-V}$ as the information is lost blue-ward of the break. 
Likewise there must be data points, or limits, blue-ward of the break so as to constrain redshift.

Plotting the various fit parameters for individual bursts against each other we see that there may be a dependency between host extinction and spectral index (as an example see GRB\,050814, Figure\,\ref{050814_scatter}): increasing the spectral index of the source, reduces the need for host extinction at a given redshift.  
There may also be a dependency between the redshift and host extinction, as a high redshift shifts highly extincted light into the observed range, though this effect is not obvious in our results.

These dependencies should be kept in mind when using the best fit parameters but because in our Monte Carlo error analysis no one parameter is fixed, as it would be in a normal $\Delta\chi^{2}$ analysis, these dependencies do not affect the quoted uncertainties.
From Figure\,\ref{050814_scatter} one can also see the hard limits set on the value of the spectral index in the fitting routine for this particular SED: $0.1 < \beta < 1.7$. Limits are set, out of computational necessity, to limit the parameter space being searched by the fitting routine. These limits must be chosen carefully, after an initial trial fit, so that they include a realistic range of parameters; wide enough to include the best fit parameters of each Monte Carlo trial but no so wide that the best fit parameters will be difficult to localise in the parameter space.

\subsection{Errors}\label{grbz:errors}

\begin{figure*}[!]
 \centering 
\resizebox{\hsize}{!}{\includegraphics[angle=-90]{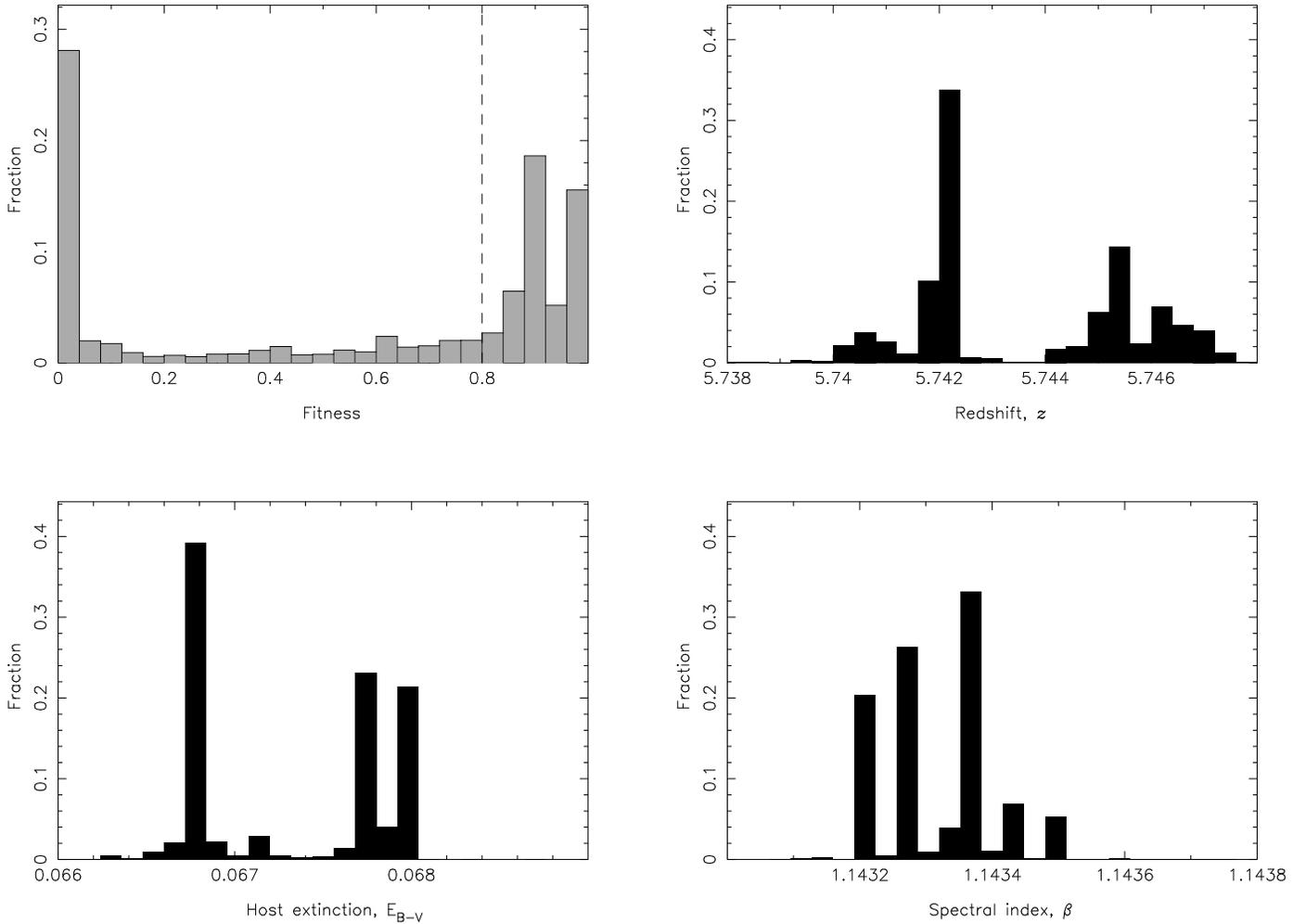} }
\caption{Distribution plots for the initial genetic algorithm fit of GRB\,050814. Redshift, host extinction and spectral index distributions are plotted for members of the optimum population with a fitness greater than 0.8.}  
 \label{census-plot}  
\end{figure*}

\begin{figure*}[!]
 \centering 
\resizebox{\hsize}{!}{\includegraphics[angle=-90]{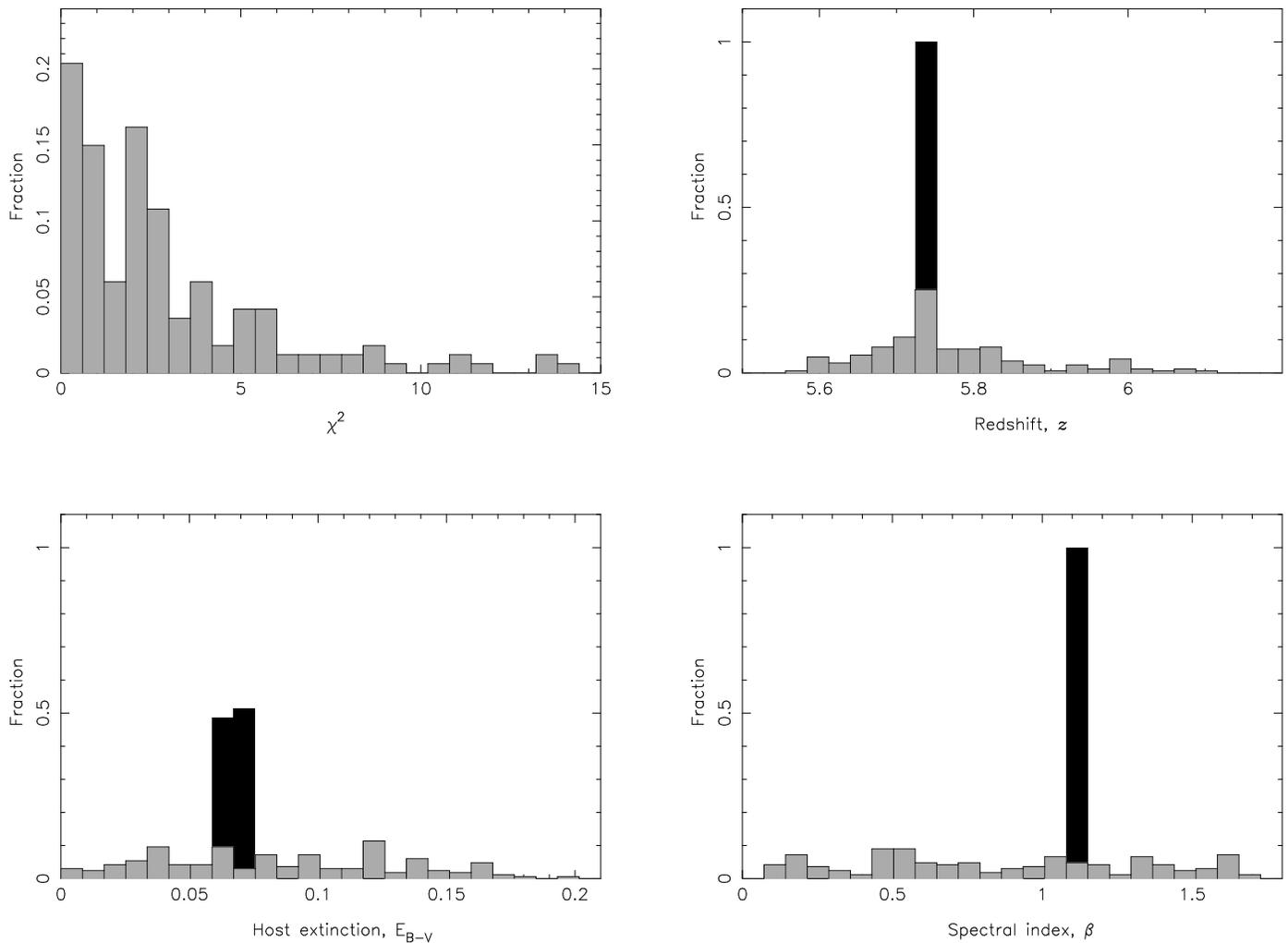} }
\caption{Distribution plots for the Monte Carlo error analysis of GRB\,050814. Redshift, host extinction and spectral index distributions are plotted for all trials of the Monte Carlo analysis, and overlaid on the re-binned results of the genetic algorithm optimum population from Figure \ref{census-plot} (in black).}  
 \label{monte-plot}  
\end{figure*}

In genetic algorithm-based optimisation, the population refers to every set of parameters that were fit in the search for the minimum $\chi^2$, though in the genetic algorithm it is `fitness' (inversely proportional to  $\chi^2$) that is maximised.  By plotting the distribution of the fitnesses of the entire population, one can see that there are many fits of poor fitness, and a number of high fitness that returned the best fit parameters. By filtering on that subset of the population that were most fit (the optimum population), one finds a distribution of the interesting parameters, the width of which is taken as an estimate of the uncertainty of the fitting mechanism. 

To illustrate, we show these distributions for one of the above bursts, GRB\,050814 (Figure\,\ref{census-plot}). We, arbitrarily though conservatively, take the optimum population to be those with a fitness greater than 0.8. This optimum population consists of two sub-populations with fitnesses centred on $\sim 0.9$ and  $\sim 1.0$, and these two populations can also be seen in the distribution plots of redshift and host extinction. From these plots, we can estimate the best fit parameters of GRB\,050814 to be: $z \sim 5.744 \pm 0.003$, $E_{B-V} \sim 0.067 \pm 0.001$, $\beta \sim 1.1434 \pm 0.0002$. 
The optimum population width method of uncertainty estimation as described by \citeN{mokiem2005:A&A441} is significantly quicker than a Monte Carlo analysis as it requires only one trial of the fitting mechanism, however, if compared to the parameters for this burst in Table \ref{result-table}, the errors are also significantly underestimated. They are more likely an estimate of the accuracy of the fitting mechanism in finding the nominal minimum for a given set of data points, than an estimate of the uncertainties due to the errors on the data points.

The difference in error estimates between the optimum population width method and the Monte Carlo analysis is clearly visible in Figure \ref{monte-plot} where we plot the distributions for all trials of our Monte Carlo error analysis of this burst, overlaid on the re-binned distributions of the optimum population of the first trial (Figure \ref{census-plot}). 
Though many of these trials return unacceptable $\chi^2$ (plotted as opposed to the fitness of the population width), they are the best fits for the given data and the average parameters and errors do not show any deviation when the unacceptable trials are removed. The distributions of the interesting parameters are clearly not Gaussian and hence the usual method of quoting sigma errors is not valid. We do quote the uncertainties as symmetric 68\% confidence intervals but it is more correct to judge the errors from these distribution plots. One can see that the host extinction and spectral index of this particular burst have almost flat distributions so it would be more proper to give each as a range as opposed to central values with errors: $E_{B-V} \lesssim 0.17$, $0.1  \lesssim \beta \lesssim 1.7$ (which corresponds to the hard limits imposed on $\beta$). Likewise the redshift would be better described with non-symmetric, though still non-Gaussian, errors as $\sim 5.7^{+0.3}_{-0.1}$.

We find similar situations, regarding the underestimate of errors by the optimum population width method and the non-Gaussian Monte Carlo errors, for the other bursts in this sample as well as a number of synthesised data sets we tested our code on. The 68\% confidence interval of the Monte Carlo error analysis should hence be treated as an approximate estimate of error.

%%**************************************************
%% Section 6: Conclusion
%% **************************************************

\section{Conclusion}\label{grbz:conclusions}

We have developed a method whereby early time photometric nIR, optical and UV data of the afterglow can be used to estimate the photometric redshift of GRBs.  This has advantages over the more usual methods of spectroscopic redshift determination, in that photometric observations do not require the source to be as bright, and over photometric estimates of the host galaxy as the afterglow is frequently much brighter than the host.
In this implementation, we assume that GRB afterglow spectra are well represented by a power-law, and model the effects of absorption due to the Lyman forest and host extinction. We use a genetic algorithm-based routine to simultaneously fit the parameters of interest, relatively unbiased by assumptions of the spectral index or host extinction.  We introduce new nIR data for GRB\,990510, which we have fitted along with the previously published data, to give new estimates of the host extinction and spectral index. We offer a new photometric redshift for GRB\,050814, slightly higher than that previously suggested. 

The Monte Carlo error analysis, though computationally time consuming, is required as the method of error estimate based on the optimum population width of the genetic algorithm underestimates errors significantly. As the distribution of the best fit parameters obtained via Monte Carlo analysis are not Gaussian, caution is required when interpreting the nominal 68\% errors of the average parameters.

%%**************************************************
%% Acknowledgements
%%**************************************************

\begin{acknowledgements}  
We acknowledge the assistance of M.R. Mokiem in the implementation of the genetic algorithm code, and for constructive conversations. We thank the referee, P. Jakobsson, and L. Kaper for useful comments. 
PAC \& RAMJW acknowledge the support of NWO under grant 639.043.302.
RLCS \& KW acknowledge support from STFC.
AJvdH is supported by an appointment to the NASA Postdoctoral Program at the NSSTC, 
administered by Oak Ridge Associated Universities through a contract with NASA.
\end{acknowledgements}

%\bibliographystyle{aa}
%\bibliography{grbzbib}

\end{document}